\documentclass[final,english]{bullsrsl}[2022/06/15]
\usepackage[latin1]{inputenc}
\usepackage[T1]{fontenc}
\usepackage{natbib} 
\usepackage{graphicx}
\usepackage{hyperref}
\usepackage{amsmath}

\begin{document}
\title{Exploring the AGN Accretion Disks using Continuum Reverberation Mapping}

\author[affil={1,2}, corresponding]{Vivek Kumar}{Jha}
\author[affil={3}]{Ravi}{Joshi}
\author[affil={4}]{Jayesh}{Saraswat}
\author[affil={5}]{Hum}{Chand}
\author[affil={3}]{Sudhanshu}{Barway}
\author[affil={6}]{Amit Kumar}{Mandal}

\affiliation[1]{Aryabhatta Research Institute of observational sciences, Nainital,  { \it 263001}; India}
\affiliation[2]{Deen Dayal Upadhyaya Gorakhpur University, Gorakhpur, { \it 273009}; India}

\affiliation[3]{Indian Institute of Astrophysics, Koramangala, Bangalore,  { \it 560034}; India}

\affiliation[4]{Department of Physics, Savitribai Phule Pune University, Pune, { \it 411007}; India}

\affiliation[5]{Department of Physics and Astronomical Sciences, Central University of Himachal Pradesh, Dharamshala, { \it 176215}; India}
\affiliation[6]{Department of Physics and Astronomy, Seoul National University, Seoul {\it 08826}, Republic of Korea}
\correspondance{vivekjha.aries@gmail.com}
\date {20th May 2023}

\maketitle
\begin{abstract}

In the innermost regions of Active Galactic Nuclei (AGN), matter is understood to be flowing onto the Supermassive black hole (SMBH), which forms an accretion disk. This disk is responsible for the optical/UV continuum emission observed in the spectra of AGN. Reverberation Mapping of the accretion disk using multiple bands can yield the structure of the disk. The emission is expected to be of the black body type peaking at different wavelengths. Hence, depending on the temperature of the disk, continuous, simultaneous monitoring in multiple wavelength ranges to cover hotter inner regions and cooler outer regions can yield the structure and temperature profile of the accretion disk itself. In this study, we present initial results from our accretion disk reverberation mapping campaign targeting AGN with Super High Eddington Accreting Black Holes (SEAMBH). Our analysis on one of the sources- IRAS 04416+1215; based on the broadband observations using the Growth India telescope (GIT), reveals that the size of the accretion disk for this source, calculated by cross-correlating the continuum light curves is larger than expected from the theoretical model. We fit the light curves directly using the thin disk model available in {\sc javelin} and find that the disk sizes are approximately 4 times larger than expected from the Shakura Sunyaev (SS) disk model. Further studies are needed to understand better the structure and physics of AGN accretion disks and their role in the evolution of galaxies.
\end{abstract}


\section{Introduction}

Active galactic nuclei (AGN) are among the most luminous and energetic objects in the universe. They are powered by the accretion of matter onto supermassive black holes at the centres of galaxies \citep{1964ApJ...140..796S}. The accretion process occurs in a disk-like structure known as an accretion disk. These disks are composed of gas and dust that orbit the black hole, gradually spiralling inward and releasing energy in the form of radiation. Understanding the properties of these disks, including their size, is crucial for understanding the behaviour of AGN and their role in the evolution of galaxies \citep{Kormendy2013}.

Reverberation mapping is a powerful technique for studying the inner structure of AGN \citep{Bahcall, Blandford1982, 2021iSci...24j2557C}. It involves measuring the time delay between variations in the continuum emission from different regions. This time delay provides a measure of the continuum emitting region. By analyzing these time delays for multiple AGNs, it is possible to measure the mass of the black hole residing at the centre \citep{2020ApJ...903..112D}. Reverberation mapping measurements have also yielded a relation between the luminosity of the AGN and the size of the Broad Line Region (BLR), the so-called R-L relation \citep{Kaspi2002,Bentz2009, Du2016}. Accretion disk reverberation mapping, which used the multi wavelength light curves to measure the lags between the various regions of the accretion disk, has been used to infer the disk sizes for a variety of objects, \citep[][etc.]{Edelson2015, Starkey2017, HernandezSantisteban2020, Kara2021}. These studies measured the inter-band lags from X-ray to the optical IR wavelengths to generate a complete profile of the accretion disk. However, only a handful of objects have such intensive measurements. Ground-based studies have also proven successful in constraining the accretion disk sizes for about a hundred AGN, although covering a smaller wavelength range \citep{Jiang2017, Mudd2018, Homayouni2020,jha2022, guo2022}.

The theoretical SS disk model has been widely used to describe the accretion disks \citep{Shakura1973}. However, results from observations have yielded that the size of the accretion disk in AGN is larger than the expectations from this model \citep{Starkey2017}. This implies that either the SS disk model does not hold for these objects or additional complexities are involved. The disk sizes being larger than predicted by standard models also imply that some key ingredients in the classical thin disk theory may need to be added.

An interesting subset of AGN are the objects accreting at super Eddington rates \citep[see][]{2013PhRvL.110h1301W}. These objects have been observed to be accreting at many times the Eddington Accretion rates. Through reverberation mapping campaigns, the BLR sizes for these objects are significantly smaller than the empirical RL relation observed in other AGN studies \citep{Du2016}. How their accretion disk sizes scale with respect to the other AGNs remains to be seen. In order to study the accretion disk structure in these objects, we are carrying out a disk reverberation mapping campaign in the optical wavelength.

In this work, we present initial results from the accretion disk reverberation mapping campaign of AGN, which we perform using data from the GROWTH India Telescope (GIT). This paper is structured as follows; Section \ref{section2} presents the sample being used for this study and the details of the observations. Section \ref{section3} presents the methods being used for our study, while Section \ref{section4} presents the initial results for one of the sources. We present the discussion in Section \ref{section5} followed by conclusions in Section \ref{section6}.

\begin{figure}
    \centering
    \includegraphics[height=10.5cm, width=10cm]{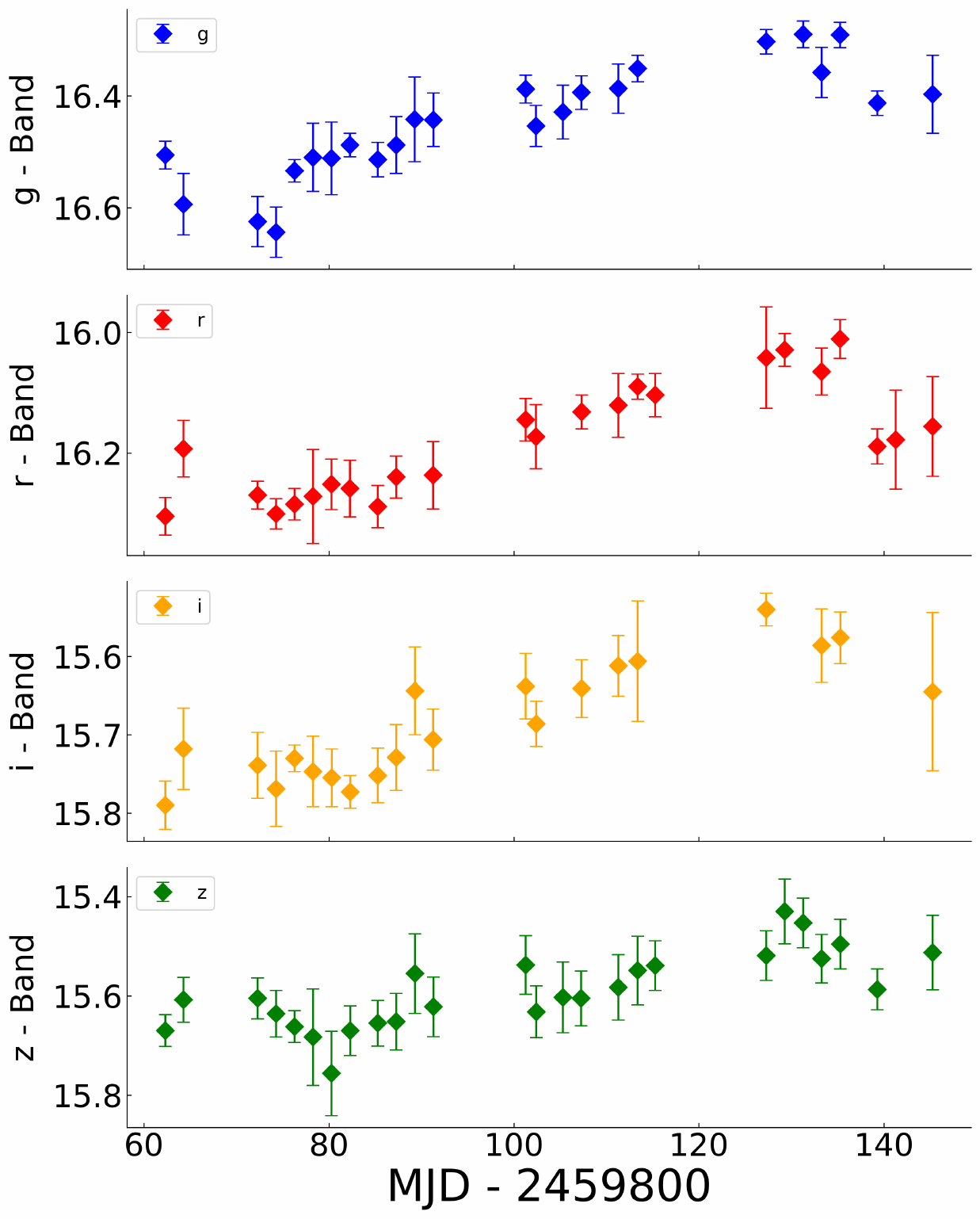}
    \caption{The g, r, i, and z band light curves for IRAS 04416+1215 were obtained using the Growth India Telescope (GIT). }
    \label{fig:my_label}
\end{figure}

\section{The Sample and Observations}
\label{section2}


We have compiled a sample of 18 AGNs with high accretion rates obtained from \citet{Du2015}. The SMBH masses for these AGNs are well constrained through the reverberation mapping studies. The BLR sizes for these sources are significantly lower than the radius luminosity relation, which has been observed to be very tight for the other AGNs with RM measurements. Whether this peculiar behaviour is seen in the accretion disk measurements of such AGN is unexplored at the moment. To map the accretion disk of these AGNs, we have started a large program titled: `Investigating the central parsec regions around supermassive black holes' (INTERVAL). In this campaign,  we are performing multi-band (u, g, r, i, and z) monitoring of AGNs using a 70cm GROWTH India telescope (GIT) and 50cm telescope at the Indian Astronomical Observatory (IAO) in Hanle, Ladakh.

The 70cm GIT  operates in robotic mode through a queue-based observation schedule. Calibration frames are taken every night, and the automatic pipeline for photometry yields the aperture and the Point Spread Function (PSF) based magnitudes for the objects \citep{growth2022}. The observations for four of the AGNs in our sample have been completed with GIT while we continue the observations for the remaining sources from our sample. The data is reduced and calibrated using standard procedures to produce light curves for each object in each band using PSF photometry. We then perform cross-correlation analysis to measure the time delay between variations in the continuum emission arising from the accretion disk.

\section {Methods}
\label{section3}

We aim to estimate the disk sizes in the AGN by measuring the interband lags between the continuum emission from different regions of the accretion disk represented by the u, g, r, i, and z band light curves. We employ two methods to derive the interband lags: {\sc javelin} and ICCF. We also use the {\sc javelin} thin disk model developed by \citet{Mudd2018} to fit the light curves directly to a disk model.

{\sc javelin} is a method that models the AGN variability as a Damped Random Walk (DRW) and employs a Bayesian approach to infer the posterior distribution of the lags and their uncertainties \citep{Zu2011}. This method has been demonstrated to be accurate and reliable as compared to the other methods being used \citep{Li2019}. ICCF is another method that computes the cross-correlation function of the light curves and identifies the peak of the function as the lag \citep{Peterson1998}. This method has been extensively applied in reverberation mapping studies and performs well when the data quality is high \citep{Sun_pyccf}.

We compare the outcomes of both methods to assess their consistency and robustness. We also conduct various tests to evaluate the validity of our measurements and respective errors. We run Markov Chain Monte Carlo (MCMC) iterations in {\sc javelin} with the parameters: nwalkers 1000, nburn=500, and nchain=1000. These parameters are sufficient, and increasing them does not affect the results. Similarly, for the ICCF method, we run 5000 iterations of Random Subset Sampling and flux Randomization in ICCF. This enables us to obtain the lags between the light curves and their associated uncertainties.

\begin{table}

\caption{\bf The light curve statistics for IRAS 04416+1215 being used for this study.}

    \centering
    \begin{tabular}{llcccc}
        \hline
        Bands & Date range (MJD) & Points & Median PSF Magnitude & Error & Average cadence (days)\\ \hline
        g & 2459835 - 2459945 & 24 & 16.44 & 0.04 & 4.5\\ 
        r & 2459835 - 2459945 & 24 & 16.18 & 0.04 & 4.5 \\ 
        i & 2459835 - 2459945 & 21 & 15.71 & 0.04 & 5.2\\ 
        z & 2459835 - 2459945 & 26 & 15.60 & 0.05 & 4.2 \\ 
        \hline
    \end{tabular}
   
\end{table}

\section{ Results:}
\label{section4}

\begin{figure}
    \centering
    \includegraphics[height=6cm, width=12cm]{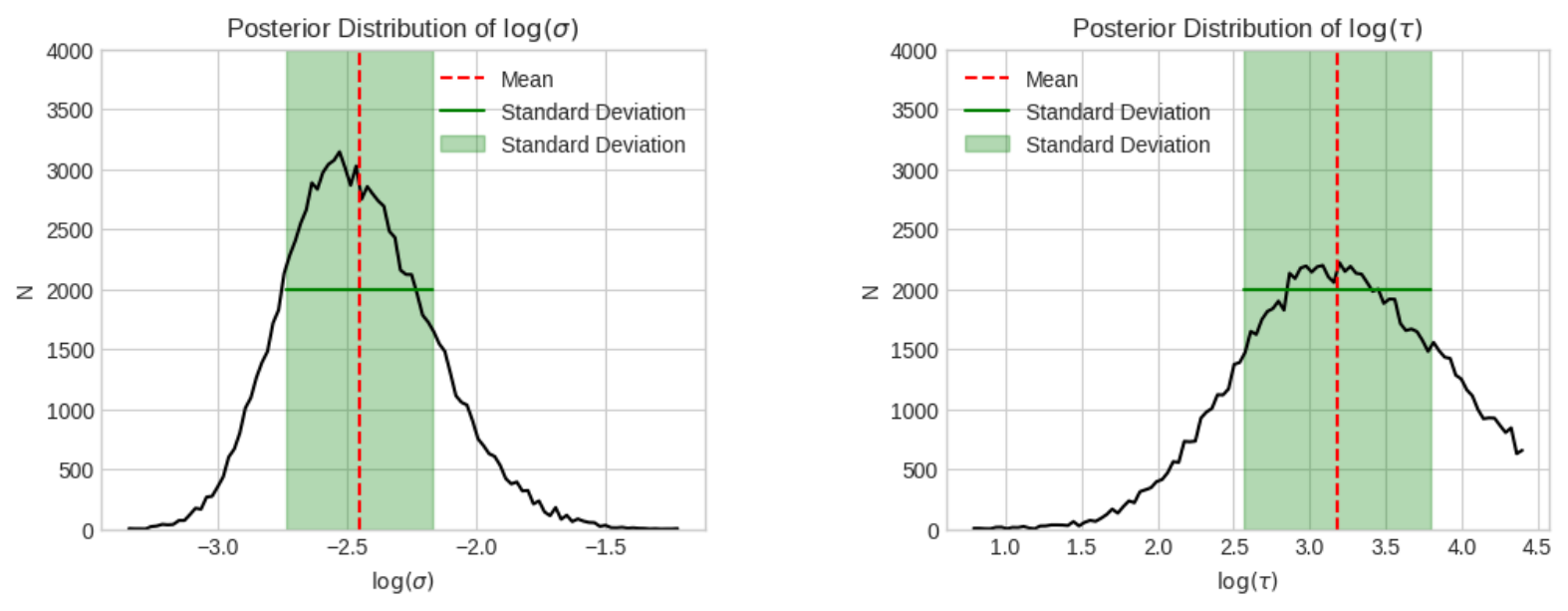}
    \caption{Posterior distribution for the logarithms of the amplitude of variability ($\sigma$) and the damping timescale ($\tau$) obtained for the driving continuum light curve, which we assume to be the g-band light curve.}
    \label{fig:my_label2}
\end{figure}

The observations for the first 4 sources in our sample, namely IRAS 4416+1215, Mrk 382, Mrk 42 and Mrk 1044, are complete, and in this work, we report on the initial results obtained for IRAS 04416+1215. It is located at a redshift of 0.0889 and categorized as a hyper Eddington source \citep{tortosa2022}.  For this source, the observations are available for a total of 27-30 points in each band. The average photometric error is about 0.04 magnitudes in all the bands. We reject the outlier points whose magnitude is greater or less than 2 magnitudes from the mean magnitude.   This reduces about 2-4 data points in each band. We use the PSF light curves for our analysis of this source. We note that the source is not prominent in the u band. Hence we use only the 4 band light curves $-$, namely the g, r, i, and z bands, for our analysis. The object appeared to be variable in all 4 bands (see Figure \ref{fig:my_label}). The inter-band lags are measured with respect to the shortest wavelength available, which happens to be the g-band.

First, we estimate the lags using {\sc javelin} \citep{Zu2011}. It models the driving light curve as a DRW process, building a posterior distribution for the variability amplitude ($\sigma$) and the damping timescale ($\tau$) (see Figure \ref{fig:my_label2}). It then shifts, scales, and smooths the light curve to generate a responding light curve and builds a distribution of the lag between the two light curves. We obtained a lag of $0.09 ^{+1.05}_{-1.07}$ days between the g-r bands, $6.93 ^{+3.69}_{-9.11}$ days between the g-i bands and $6.26 ^{+5.23}_{-5.78}$ days between the g-z bands. Noticeably, the g-r band lag is larger than the g-i band lags, although the uncertainties are significantly higher at the 1 $\sigma$ level.

\begin{figure}
    \centering
    \includegraphics[height=4.5cm, width=17cm]{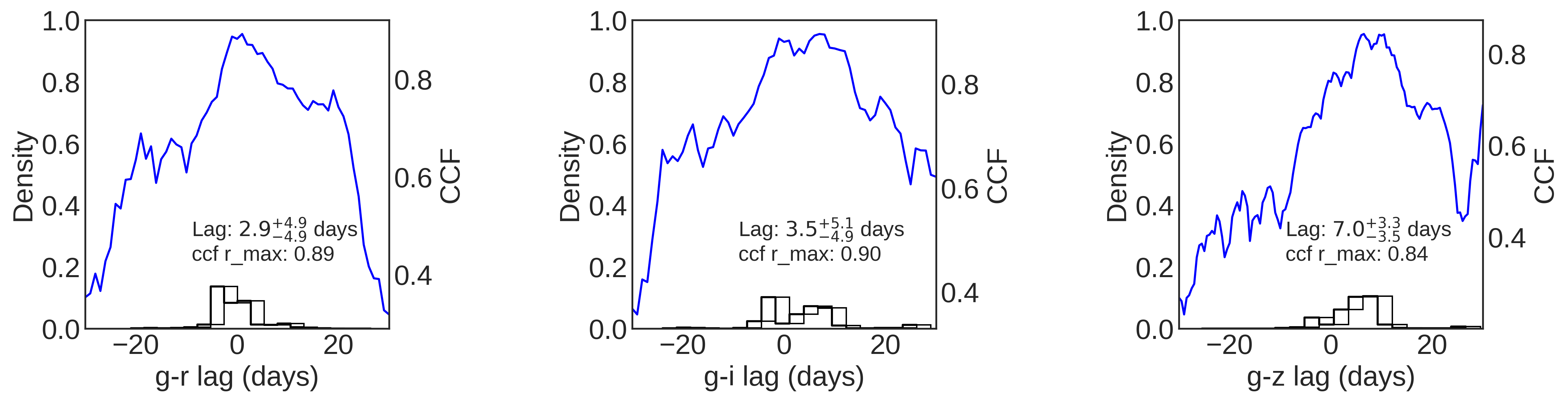}
    \caption{The Cross Correlation Posterior Distribution (CCPD) along with the correlation coefficient for IRAS 04416+1215 for the g-r, g-i, and g-z bands are shown here.}
    \label{fig:my_label3}
\end{figure}

\begin{table}
\caption{\bf Estimated lags using {\sc javelin} and ICCF (in days) for IRAS 04416+1215.}
\vspace{0.3cm}
    \centering
    \begin{tabular}{llcc}
    \hline
        Bands & {\sc javelin} lags  &  ICCF Lags \\ \hline

        g - r & $0.09 ^{+1.05}_{-1.07}$  & $2.9 ^{+4.9}_{-4.9}$ \\
        
        g - i & $6.93 ^{+3.69}_{-9.11}$ & $3.5 ^{+5.1}_{-4.9}$ \\
        
        g - z & $6.26 ^{+5.23}_{-5.78}$  & $7.0 ^{+3.3}_{-3.5}$ \\
        \hline
    \end{tabular}

\end{table}

We also implemented the ICCF method, a widely used method for the lag measurements \citep{Peterson1998}. It does not assume an underlying model, but rather it interpolates between the gaps and then cross-correlates the individual points to generate a distribution for the cross-correlation function. The uncertainties are estimated using flux randomization (FR) and random subset selection (RSS). We use the centroid of the CCF to estimate the lag. We obtain a lag of  $2.9 ^{+4.9}_{-4.9}$   days for g-r band, $3.5 ^{+5.1}_{-4.9}$ days for g-i band and $7.0 ^{+3.3}_{-3.5}$ days for g-z bands (see Figure \ref{fig:my_label3}). The increasing lags with the increase in wavelength are seen using the ICCF method. We find that the obtained lags through both methods are slightly different with higher uncertainties, especially with the measurement using {\sc javelin}.

We notice that the uncertainties in the lag estimates are quite large using both {\sc javelin} and the ICCF methods. One of the primary reasons could be the cadence of the light curves. The median cadence that we achieve is about 4-5 days, which makes it difficult to recover shorter lags. For other sources in our sample, we aim to get denser sampling so as to recover lags with better precision.

Taking the advantage of our multi-band data set, we fit the light curves using the {\sc javelin} thin disk model. The g-band light curve is assumed to be the driving light curve, while the r, i, and z-band light curves are responding light curves. The disk model gives us the disk size at the driving wavelength, which is the g-band rest wavelength. We fixed the $\beta$ parameter to 1.33 while calculating the disk size, which implies a SS disk. The disk size at the g-band rest wavelength is estimated to be $2.96^{+3.28}_{-2.03}$ light days.

We also calculate the disk size $R$ size at the g-band rest wavelength  $\lambda_0$, using the equation in \citet{Mudd2018}, based on  the SS disk model as:

\begin{equation}
    R_{\lambda0} = 9.7 \times 10^{15} \left( \frac{\lambda_0}{\mu \text{m}} \right)^\beta \left( \frac{M}{10^9 \textup{M}_\odot} \right)^{2/3} \left( \frac{L}{\eta L_\text{E}} \right)^{1/3} \text{cm}
    \label{eq:Thin Disk Eqn}
\end{equation}

Where ${\lambda_0}$ is the wavelength at which the disk size is estimated, $\beta = 4/3$, M is the SMBH mass in units of Solar mass (${M}_\odot$), and  L is the luminosity obtained from \citep{Du2015}, $ L_\text{E}$ is the Eddington luminosity. We take the efficiency parameter $\eta$ as 0.1, as has been used in previous works.

\begin{figure}
    \centering
    \includegraphics[height=7cm, width=11cm]{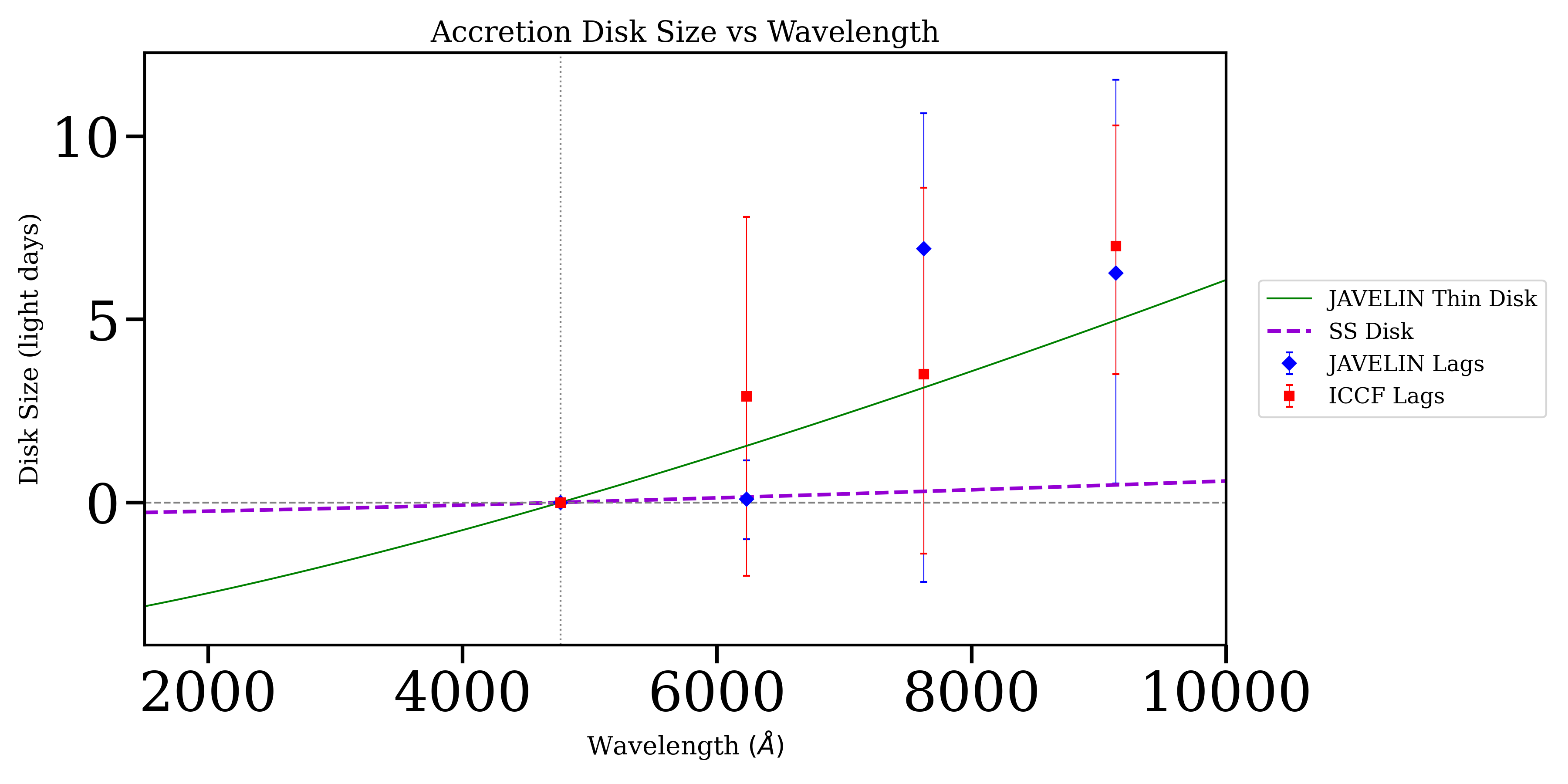}
    \caption{The relation between the accretion disk size and the wavelength for IRAS 04416+1215. We use the g-band rest wavelength as the reference wavelength for estimating the disk sizes. The green solid line shows the extrapolation on the {\sc javelin} thin disk model. The blue points show the lags obtained using  {\sc javelin}, while the red points show the lags obtained using ICCF. The dashed purple line shows the standard SS disk model prediction for this source. }
    \label{fig:my_label4}
\end{figure}

We find that the disk size obtained using the {\sc javelin} thin disk model is 4 times larger than what we would obtain based on the theoretical prediction of the SS disk as given in equation \ref{eq:Thin Disk Eqn} . If we plot the thin disk model, we find that the calculated lags lie near to the curve generated by this model, implying that the disk is following the 4/3 scaling relation, but the size of the disk is larger than the prediction for the SS disk (see Figure \ref{fig:my_label4}).

\section {Discussion}
\label{section5}

In this study, we present initial results from the `INTERVAL' campaign measuring the accretion disk sizes for SEAMBH AGN using the GIT. For one of the sources, we measure the inter-band time delays using both the {\sc javelin} and ICCF methods. Our analysis reveals that the size of the accretion disks for this object is larger than expected from the theoretical SS disk model. Our findings are consistent with recent studies that have found that AGN accretion disk sizes are generally larger than predicted by standard models \citep{Starkey2017, Fausnaugh2018, Edelson2019}. Recent studies using optical data from the Zwicky Transient Facility have also found that the accretion disk sizes for AGN are multiple times larger than predicted by the standard SS disk model \citep{jha2022, weijian2022, guo2022}. 

To address this discrepancy, it has been proposed that the larger-than-expected accretion disk sizes found in AGN studies are primarily due to the neglect of reddening \citep{Gaskell2017}. They estimated that neglecting internal extinction leads to an underestimate of the luminosity at 1200\AA{}  by a factor of seven, and therefore the size scale of the accretion disk has been underestimated by a factor of about 2.6. This is similar to the accretion disk size discrepancy found in other studies and supports the proposal that internal reddening plays a significant role in explaining the discrepancy. We aim to estimate the reddening for our sources, to test whether this phenomenon affects the lag estimates. The contribution of the Diffuse Continuum (DC) to the continuum light curves has also been observed \citep[see][]{Chelouche2019}, which can have a direct consequence of increasing the interband continuum lags on account of the contribution from the BLR.

Another possibility is that the accretion disks in AGN are not geometrically flat but instead have a slim \citep{Narayan1995} or even a clumpy structure, which would not agree well with the SS disk model. Analytical modelling of the accretion disk will be very helpful in resolving this discrepancy. The results suggest that there are some key ingredients in the classical thin disk theory that need to be added. Further studies are needed to investigate this possibility and to better understand the structure and physics of AGN accretion disks.

While the BLR sizes for SEAMBH AGN are estimated to be lower than estimated from the R-L relation, we also want to explore whether this phenomenon is true for the accretion disk sizes. However, our result for one of the SEAMBH sources, namely IRAS 04416+1215, shows that the disk size is about 4 times larger than the SS prediction, a phenomenon observed for the other AGNs as well. We continue to monitor these AGNs in order to better understand their accretion disk structure.

\section {Conclusions}
\label{section6}

We present results from our accretion disk reverberation mapping campaign targeting AGN with Super High Accretion Rates. Initial results for one of the sources, IRAS 04416+1215, identified as a hyper Eddington accreting source, are presented. Here are the conclusions from this work:

\begin{enumerate}

\item The interband continuum lags analysis for IRAS 04416+1215 reveals that the size of the accretion disk is larger than expected from the standard SS disk model.

     \item We fit the light curves directly using the thin disk model available in the {\sc javelin} package, which reveals the disk sizes to be about 4 times larger than the prediction of the theoretical SS disk model.
  
\item Our findings are consistent with recent studies that have found that AGN accretion disk sizes are generally larger than predicted by the SS disk model. These results suggest that there can be some key ingredients in the classical thin disk theory that need to be added.

\item For further understanding, we continue to monitor the sample of SEAMBH AGN in order to get the accretion disk size estimates based on continuum reverberation mapping. Such studies are needed to understand better the structure and physics of AGN accretion disks and their role in the evolution of galaxies.
\end{enumerate}


\begin{acknowledgments}
 This research is part of DST-SERB project grant no. SRG/2022/001785. RJ acknowledge the financial support provided by DST-SERG for this work.
We would like to express our gratitude to the GROWTH India telescope team for their support and assistance in obtaining the data used in this study. We also thank ARIES for providing the facilities that made this work possible. Finally, we are grateful to the BINA organizers for giving us the opportunity to present our work at their conference. This research would not have been possible without the contributions of these individuals and organizations.
\end{acknowledgments}

\begin{furtherinformation}

\begin{authorcontributions}
The author contribution is as follows;\\ VKJ: Conceptualization, Writing Original Draft, RJ: Writing $-$ Review \& Editing, Resources, Project Administration,  JS: Writing $-$  Review \& Editing, Formal Analysis,  HC:  Writing $-$  Review \& Editing, Supervision, SB: Writing $-$  Review \& Editing, Resources, Project Administration,  AKM:  Review \& Editing, Resources, Project Administration.
\end{authorcontributions}

\begin{conflictsofinterest}
The authors declare no conflict of interest.
\end{conflictsofinterest}

\end{furtherinformation}

\bibliographystyle{aasjournal}

\bibliography{main}

\begin{thebibliography}{}
\expandafter\ifx\csname natexlab\endcsname\relax\def\natexlab#1{#1}\fi
\providecommand{\url}[1]{\href{#1}{#1}}
\providecommand{\dodoi}[1]{doi:~\href{http://doi.org/#1}{\nolinkurl{#1}}}
\providecommand{\doeprint}[1]{\href{http://ascl.net/#1}{\nolinkurl{http://ascl.net/#1}}}
\providecommand{\doarXiv}[1]{\href{https://arxiv.org/abs/#1}{\nolinkurl{https://arxiv.org/abs/#1}}}

\bibitem[{Bahcall {et~al.}(1972)Bahcall, Kozlovsky, \& Salpeter}]{Bahcall}
Bahcall, J.~N., Kozlovsky, B.-Z., \& Salpeter, E.~E. 1972, The Astrophysical
  Journal, 171, 467, \dodoi{10.1086/151300}

\bibitem[{Bentz {et~al.}(2009)Bentz, Walsh, Barth, Baliber, Bennert, Canalizo,
  Filippenko, Ganeshalingam, Gates, Greene, Hidas, Hiner, Lee, Li, Malkan,
  Minezaki, Sakata, Serduke, Silverman, Steele, Stern, Street, Thornton, Treu,
  Wang, Woo, \& Yoshii}]{Bentz2009}
Bentz, M.~C., Walsh, J.~L., Barth, A.~J., {et~al.} 2009, The Astrophysical
  Journal, 705, 199, \dodoi{10.1088/0004-637X/705/1/199}

\bibitem[{Blandford \& McKee(1982)}]{Blandford1982}
Blandford, R.~D., \& McKee, C.~F. 1982, The Astrophysical Journal, 255, 419,
  \dodoi{10.1086/159843}

\bibitem[{{Cackett} {et~al.}(2021){Cackett}, {Bentz}, \&
  {Kara}}]{2021iSci...24j2557C}
{Cackett}, E.~M., {Bentz}, M.~C., \& {Kara}, E. 2021, iScience, 24, 102557,
  \dodoi{10.1016/j.isci.2021.102557}

\bibitem[{Chelouche {et~al.}(2019)Chelouche, {Pozo Nu{\~{n}}ez}, \&
  Kaspi}]{Chelouche2019}
Chelouche, D., {Pozo Nu{\~{n}}ez}, F., \& Kaspi, S. 2019, Nature Astronomy, 3,
  251, \dodoi{10.1038/s41550-018-0659-x}

\bibitem[{{Dalla Bont{\`a}} {et~al.}(2020){Dalla Bont{\`a}}, {Peterson},
  {Bentz}, {Brandt}, {Ciroi}, {De Rosa}, {Fonseca Alvarez}, {Grier}, {Hall},
  {Hern{\'a}ndez Santisteban}, {Ho}, {Homayouni}, {Horne}, {Kochanek}, {Li},
  {Morelli}, {Pizzella}, {Pogge}, {Schneider}, {Shen}, {Trump}, \&
  {Vestergaard}}]{2020ApJ...903..112D}
{Dalla Bont{\`a}}, E., {Peterson}, B.~M., {Bentz}, M.~C., {et~al.} 2020, \apj,
  903, 112, \dodoi{10.3847/1538-4357/abbc1c}

\bibitem[{Du {et~al.}(2015)Du, Hu, Lu, Huang, Cheng, Qiu, Li, Zhang, Fan, Bai,
  Bian, Yuan, Kaspi, Ho, Netzer, \& Wang}]{Du2015}
Du, P., Hu, C., Lu, K.-X., {et~al.} 2015, The Astrophysical Journal, 806, 22,
  \dodoi{10.1088/0004-637X/806/1/22}

\bibitem[{Du {et~al.}(2016)Du, Lu, Zhang, Huang, Wang, Hu, Qiu, Li, Fan, Fang,
  Bai, Bian, Yuan, Ho, \& Wang}]{Du2016}
Du, P., Lu, K.-X., Zhang, Z.-X., {et~al.} 2016, The Astrophysical Journal, 825,
  126, \dodoi{10.3847/0004-637x/825/2/126}

\bibitem[{Edelson {et~al.}(2015)Edelson, Gelbord, Horne, McHardy, Peterson,
  Ar{\'{e}}valo, Breeveld, Rosa, Evans, Goad, Kriss, Brandt, Gehrels, Grupe,
  Kennea, Kochanek, Nousek, Papadakis, Siegel, Starkey, Uttley, Vaughan, Young,
  Barth, Bentz, Brewer, Crenshaw, {Dalla Bont{\`{a}}}, C{\'{a}}ceres, Denney,
  Dietrich, Ely, Fausnaugh, Grier, Hall, Kaastra, Kelly, Korista, Lira, Mathur,
  Netzer, Pancoast, Pei, Pogge, Schimoia, Treu, Vestergaard, Villforth, Yan, \&
  Zu}]{Edelson2015}
Edelson, R., Gelbord, J.~M., Horne, K., {et~al.} 2015, The Astrophysical
  Journal, 806, 129, \dodoi{10.1088/0004-637X/806/1/129}

\bibitem[{Edelson {et~al.}(2019)Edelson, Gelbord, Cackett, Peterson, Horne,
  Barth, Starkey, Bentz, Brandt, Goad, Joner, Korista, Netzer, Page, Uttley,
  Vaughan, Breeveld, Cenko, Done, Evans, Fausnaugh, Ferland, Gonzalez-Buitrago,
  Gropp, Grupe, Kaastra, Kennea, Kriss, Mathur, Mehdipour, Mudd, Nousek,
  Schmidt, Vestergaard, \& Villforth}]{Edelson2019}
Edelson, R., Gelbord, J., Cackett, E., {et~al.} 2019, The Astrophysical
  Journal, 870, 123, \dodoi{10.3847/1538-4357/aaf3b4}

\bibitem[{Fausnaugh {et~al.}(2018)Fausnaugh, Starkey, Horne, Kochanek,
  Peterson, Bentz, Denney, Grier, Grupe, Pogge, {De Rosa}, Adams, Barth,
  Beatty, Bhattacharjee, Borman, Boroson, Bottorff, Brown, Brown, Brotherton,
  Coker, Crawford, Croxall, Eftekharzadeh, Eracleous, Joner, Henderson,
  Holoien, Hutchison, Kaspi, Kim, King, Li, Lochhaas, Ma, MacInnis,
  Manne-Nicholas, Mason, Montuori, Mosquera, Mudd, Musso, Nazarov, Nguyen,
  Okhmat, Onken, Ou-Yang, Pancoast, Pei, Penny, Poleski, Rafter,
  Romero-Colmenero, Runnoe, Sand, Schimoia, Sergeev, Shappee, Simonian, Somers,
  Spencer, Stevens, Tayar, Treu, Valenti, {Van Saders}, Villanueva, Villforth,
  Weiss, Winkler, \& Zhu}]{Fausnaugh2018}
Fausnaugh, M.~M., Starkey, D.~A., Horne, K., {et~al.} 2018, The Astrophysical
  Journal, 854, 107, \dodoi{10.3847/1538-4357/aaaa2b}

\bibitem[{{Gaskell}(2017)}]{Gaskell2017}
{Gaskell}, C.~M. 2017, \mnras, 467, 226, \dodoi{10.1093/mnras/stx094}

\bibitem[{{Guo} {et~al.}(2022{\natexlab{a}}){Guo}, {Barth}, \&
  {Wang}}]{guo2022}
{Guo}, H., {Barth}, A.~J., \& {Wang}, S. 2022{\natexlab{a}}, \apj, 940, 20,
  \dodoi{10.3847/1538-4357/ac96ec}

\bibitem[{{Guo} {et~al.}(2022{\natexlab{b}}){Guo}, {Li}, {Zhang}, {Ho}, \&
  {Wang}}]{weijian2022}
{Guo}, W.-J., {Li}, Y.-R., {Zhang}, Z.-X., {Ho}, L.~C., \& {Wang}, J.-M.
  2022{\natexlab{b}}, \apj, 929, 19, \dodoi{10.3847/1538-4357/ac4e84}

\bibitem[{{Hern{\'{a}}ndez Santisteban} {et~al.}(2020){Hern{\'{a}}ndez
  Santisteban}, Edelson, Horne, Gelbord, Barth, Cackett, Goad, Netzer, Starkey,
  Uttley, Brandt, Korista, Lohfink, Onken, Page, Siegel, Vestergaard, Bisogni,
  Breeveld, Cenko, {Dalla Bont{\`{a}}}, Evans, Ferland, Gonzalez-Buitrago,
  Grupe, Joner, Kriss, LaPorte, Mathur, Marshall, Mehdipour, Mudd, Peterson,
  Schmidt, Vaughan, \& Valenti}]{HernandezSantisteban2020}
{Hern{\'{a}}ndez Santisteban}, J.~V., Edelson, R., Horne, K., {et~al.} 2020,
  Monthly Notices of the Royal Astronomical Society, 498, 5399,
  \dodoi{10.1093/mnras/staa2365}

\bibitem[{Homayouni {et~al.}(2019)Homayouni, Trump, Grier, Shen, Starkey,
  Brandt, Alvarez, Hall, Horne, Kinemuchi, Li, McGreer, Sun, Ho, \&
  Schneider}]{Homayouni2020}
Homayouni, Y., Trump, J.~R., Grier, C.~J., {et~al.} 2019, The Astrophysical
  Journal, 880, 126, \dodoi{10.3847/1538-4357/ab2638}

\bibitem[{{Jha} {et~al.}(2022){Jha}, {Joshi}, {Chand}, {Wu}, {Ho}, {Rastogi},
  \& {Ma}}]{jha2022}
{Jha}, V.~K., {Joshi}, R., {Chand}, H., {et~al.} 2022, \mnras, 511, 3005,
  \dodoi{10.1093/mnras/stac109}

\bibitem[{{Jiang} {et~al.}(2017){Jiang}, {Green}, {Greene}, {Morganson},
  {Shen}, {Pancoast}, {MacLeod}, {Anderson}, {Brandt}, {Grier}, {Rix}, {Ruan},
  {Protopapas}, {Scott}, {Burgett}, {Hodapp}, {Huber}, {Kaiser}, {Kudritzki},
  {Magnier}, {Metcalfe}, {Tonry}, {Wainscoat}, \& {Waters}}]{Jiang2017}
{Jiang}, Y.-F., {Green}, P.~J., {Greene}, J.~E., {et~al.} 2017, \apj, 836, 186,
  \dodoi{10.3847/1538-4357/aa5b91}

\bibitem[{{Kara} {et~al.}(2021){Kara}, {Mehdipour}, {Kriss}, {Cackett}, {Arav},
  {Barth}, {Byun}, {Brotherton}, {De Rosa}, {Gelbord}, {Hern{\'a}ndez
  Santisteban}, {Hu}, {Kaastra}, {Landt}, {Li}, {Miller}, {Montano},
  {Partington}, {Aceituno}, {Bai}, {Bao}, {Bentz}, {Brink}, {Chelouche},
  {Chen}, {Colmenero}, {Dalla Bont{\`a}}, {Dehghanian}, {Du}, {Edelson},
  {Ferland}, {Ferrarese}, {Fian}, {Filippenko}, {Fischer}, {Goad},
  {Gonz{\'a}lez Buitrago}, {Gorjian}, {Grier}, {Guo}, {Hall}, {Ho},
  {Homayouni}, {Horne}, {Ili{\'c}}, {Jiang}, {Joner}, {Kaspi}, {Kochanek},
  {Korista}, {Kynoch}, {Li}, {Liu}, {McHardy}, {McLane}, {Mitchell}, {Netzer},
  {Olson}, {Pogge}, {Popovi{\'c}}, {Proga}, {Storchi-Bergmann}, {Strasburger},
  {Treu}, {Vestergaard}, {Wang}, {Ward}, {Waters}, {Williams}, {Yang}, {Yao},
  {Zastrocky}, {Zhai}, \& {Zu}}]{Kara2021}
{Kara}, E., {Mehdipour}, M., {Kriss}, G.~A., {et~al.} 2021, \apj, 922, 151,
  \dodoi{10.3847/1538-4357/ac2159}

\bibitem[{Kaspi {et~al.}(2000)Kaspi, Smith, Netzer, Maoz, Jannuzi, \&
  Giveon}]{Kaspi2002}
Kaspi, S., Smith, P.~S., Netzer, H., {et~al.} 2000, The Astrophysical Journal,
  533, 631, \dodoi{10.1086/308704}

\bibitem[{{Kormendy} \& {Ho}(2013)}]{Kormendy2013}
{Kormendy}, J., \& {Ho}, L.~C. 2013, \araa, 51, 511,
  \dodoi{10.1146/annurev-astro-082708-101811}

\bibitem[{{Kumar} {et~al.}(2022){Kumar}, {Bhalerao}, {Anupama}, {Barway},
  {Basu}, {Deshmukh}, {De}, {Dutta}, {Fremling}, {Iyer}, {Jassani}, {Joharle},
  {Karambelkar}, {Khandagale}, {Krishna}, {Kulkarni}, {Mate}, {Patil},
  {Phanindra}, {Samantaray}, {Sharma}, {Sharma}, {Shenoy}, {Singh},
  {Srivastava}, {Swain}, {Waratkar}, {Angchuk}, {Dorjay}, {Dorjai}, {Gyalson},
  {Jorphail}, {Mahay}, {Norbu}, {Sharma}, {Stanzin}, {Stanzin}, \&
  {Stanzin}}]{growth2022}
{Kumar}, H., {Bhalerao}, V., {Anupama}, G.~C., {et~al.} 2022, The Astronomical
  Journal, 164, 90, \dodoi{10.3847/1538-3881/ac7bea}

\bibitem[{Li {et~al.}(2019)Li, Shen, Brandt, Grier, Hall, Ho, Homayouni, Horne,
  Schneider, Trump, \& Starkey}]{Li2019}
Li, J. I.-H., Shen, Y., Brandt, W.~N., {et~al.} 2019, The Astrophysical
  Journal, 884, 119, \dodoi{10.3847/1538-4357/ab41fb}

\bibitem[{Mudd {et~al.}(2018)Mudd, Martini, Zu, Kochanek, Peterson, Kessler,
  Davis, Hoormann, King, Lidman, Sommer, Tucker, Asorey, Hinton, Glazebrook,
  Kuehn, Lewis, Macaulay, Moeller, O'Neill, Zhang, Abbott, Abdalla, Allam,
  Banerji, Benoit-L{\'{e}}vy, Bertin, Brooks, Rosell, Carollo, Kind, Carretero,
  Cunha, D'Andrea, da~Costa, Davis, Desai, Doel, Fosalba, Garc{\'{i}}a-Bellido,
  Gaztanaga, Gerdes, Gruen, Gruendl, Gschwend, Gutierrez, Hartley, Honscheid,
  James, Kuhlmann, Kuropatkin, Lima, Maia, Marshall, McMahon, Menanteau,
  Miquel, Plazas, Romer, Sanchez, Schindler, Schubnell, Smith, Smith,
  Soares-Santos, Sobreira, Suchyta, Swanson, Tarle, Thomas, Tucker, \&
  Walker}]{Mudd2018}
Mudd, D., Martini, P., Zu, Y., {et~al.} 2018, The Astrophysical Journal, 862,
  123, \dodoi{10.3847/1538-4357/aac9bb}

\bibitem[{{Narayan} \& {Yi}(1995)}]{Narayan1995}
{Narayan}, R., \& {Yi}, I. 1995, \apj, 452, 710, \dodoi{10.1086/176343}

\bibitem[{Peterson {et~al.}(1998)Peterson, Wanders, Horne, Collier, Alexander,
  Kaspi, \& Maoz}]{Peterson1998}
Peterson, B., Wanders, I., Horne, K., {et~al.} 1998, Publications of the
  Astronomical Society of the Pacific, 110, 660, \dodoi{10.1086/316177}

\bibitem[{{Salpeter}(1964)}]{1964ApJ...140..796S}
{Salpeter}, E.~E. 1964, \apj, 140, 796, \dodoi{10.1086/147973}

\bibitem[{{Shakura} \& {Sunyaev}(1973)}]{Shakura1973}
{Shakura}, N.~I., \& {Sunyaev}, R.~A. 1973, \aap, 24, 337

\bibitem[{Starkey {et~al.}(2016)Starkey, Horne, Fausnaugh, Peterson, Bentz,
  Kochanek, Denney, Edelson, Goad, {De Rosa}, Anderson, Arevalo, Barth, Bazhaw,
  Borman, Boroson, Bottorff, Brandt, Breeveld, Cackett, Carini, Croxall,
  Crenshaw, Bonta, {De Lorenzo-Caceres}, Dietrich, Efimova, Ely, Evans,
  Filippenko, Flatland, Gehrels, Geier, Gelbord, Gonzalez, Gorjian, Grier,
  Grupe, Hall, Hicks, Horenstein, Hutchison, Im, Jensen, Joner, Jones, Kaastra,
  Kaspi, Kelly, Kennea, Kim, Kim, Klimanov, Korista, Kriss, Lee, Leonard, Lira,
  MacInnis, Manne-Nicholas, Mathur, McHardy, Montouri, Musso, Nazarov, Norris,
  Nousek, Okhmat, Pancoast, Parks, Pei, Pogge, Pott, Rafter, Rix, Saylor,
  Schimoia, Schnulle, Sergeev, Siegel, Spencer, Sung, Teems, Turner, Uttley,
  Vestergaard, Villforth, Weiss, Woo, Yan, Young, Zheng, \& Zu}]{Starkey2017}
Starkey, D., Horne, K., Fausnaugh, M.~M., {et~al.} 2016, The Astrophysical
  Journal, 835, 65, \dodoi{10.3847/1538-4357/835/1/65}

\bibitem[{{Sun} {et~al.}(2018){Sun}, {Grier}, \& {Peterson}}]{Sun_pyccf}
{Sun}, M., {Grier}, C.~J., \& {Peterson}, B.~M. 2018, {PyCCF: Python Cross
  Correlation Function for reverberation mapping studies}.
\newblock \doeprint{1805.032}

\bibitem[{{Tortosa} {et~al.}(2022){Tortosa}, {Ricci}, {Tombesi}, {Ho}, {Du},
  {Inayoshi}, {Wang}, {Shangguan}, \& {Li}}]{tortosa2022}
{Tortosa}, A., {Ricci}, C., {Tombesi}, F., {et~al.} 2022, \mnras, 509, 3599,
  \dodoi{10.1093/mnras/stab3152}

\bibitem[{{Wang} {et~al.}(2013){Wang}, {Du}, {Valls-Gabaud}, {Hu}, \&
  {Netzer}}]{2013PhRvL.110h1301W}
{Wang}, J.-M., {Du}, P., {Valls-Gabaud}, D., {Hu}, C., \& {Netzer}, H. 2013,
  \prl, 110, 081301, \dodoi{10.1103/PhysRevLett.110.081301}

\bibitem[{Zu {et~al.}(2011)Zu, Kochanek, \& Peterson}]{Zu2011}
Zu, Y., Kochanek, C.~S., \& Peterson, B.~M. 2011, The Astrophysical Journal,
  735, 80, \dodoi{10.1088/0004-637X/735/2/80}

\end{thebibliography}

\end{document}